\documentclass[twocolumn]{aastex63}
\usepackage[utf8]{inputenc}
\usepackage{amsmath}
\usepackage{amsfonts}
\usepackage{amssymb}
\usepackage{graphicx}
\usepackage{grffile}
\usepackage{hyperref}
\usepackage[caption=false]{subfig}
\usepackage{array}
\usepackage{listings}
\usepackage{comment}
\usepackage{bm}
\usepackage{slashed}
\usepackage{mathtools}
\let\tablenum\relax
\usepackage{siunitx}
\usepackage{multirow}
\usepackage{booktabs}
\usepackage[para]{threeparttable}
\usepackage{ulem}
\usepackage{xcolor}

\numberwithin{equation}{section}

\newcommand*{\rH}{{\rm H}}
\newcommand*{\rD}{{\rm D}}

\def\re{{\rm e}}
\renewcommand{\refeq}[1]{Eq.~(\ref{eq:#1})}
\newcommand{\reffig}[1]{Fig.~\ref{fig:#1}}

\DeclareSIUnit \atomicunit{a.u.}
\DeclareSIUnit \erg{erg}
\DeclareSIUnit \rydberg{Ry}
\DeclareSIUnit \epccm{\erg\per\centi\meter\cubed\per\second}
\DeclareSIUnit \clight{\text{\ensuremath{c}}}

\definecolor{RedWine}{rgb}{0.743,0,0}
\definecolor{GrassGreen}{rgb}{0.125,0.75,0.125}
\definecolor{RoyalBlue}{rgb}{0.25,0.41,0.88}
\definecolor{DarkCyan}{rgb}{0,0.5,0.5}

\begin{document}
\reportnum{KIAS-P22051}

\title{A Lower Bound on the Mass of Compact Objects from Dissipative Dark Matter}


\author{James Gurian}
\affiliation{Institute for Gravitation and the Cosmos, The Pennsylvania State University, University Park, PA 16802, USA}
\affiliation{Department of Astronomy and Astrophysics, The Pennsylvania State University, University Park, PA, 16802, USA,
}

\author{Michael Ryan}
\affiliation{Institute for Gravitation and the Cosmos, The Pennsylvania State University, University Park, PA 16802, USA}
\affiliation{Department of Physics, The Pennsylvania State University, University Park, PA, 16802, USA}

\author{Sarah Schon}
\affiliation{Institute for Gravitation and the Cosmos, The Pennsylvania State University, University Park, PA 16802, USA}
\affiliation{Department of Physics, The Pennsylvania State University, University Park, PA, 16802, USA}

\author{Donghui Jeong}
\affiliation{Institute for Gravitation and the Cosmos, The Pennsylvania State University, University Park, PA 16802, USA}
\affiliation{Department of Astronomy and Astrophysics, The Pennsylvania State University, University Park, PA, 16802, USA,
}
\affiliation{School of Physics, Korea Institute for Advanced Study (KIAS), 85 Hoegiro, Dongdaemun-gu, Seoul, 02455, Republic of Korea}

\author{Sarah Shandera}
\affiliation{Institute for Gravitation and the Cosmos, The Pennsylvania State University, University Park, PA 16802, USA}
\affiliation{Department of Physics, The Pennsylvania State University, University Park, PA, 16802, USA}

\email{jhg5248@psu.edu}

\begin{abstract}

We study the fragmentation scale of dark gas formed in dissipative dark-matter halos and show that the simple atomic-dark-matter model consistent with all current observations can create low-mass fragments that can evolve into compact objects forbidden by stellar astrophysics. We model the collapse of the dark halo's dense core by tracing the thermo-chemical evolution of a uniform-density volume element under two extreme assumptions for density evolution: hydrostatic equilibrium and pressure-free collapse. We then compute the opacity-limited minimum fragment mass from the minimum temperature achieved in these calculations. The results indicate that much of the parameter space is highly unstable to small-scale fragmentation. 
\end{abstract}

\section{Introduction}
If dark matter is more similar than not to baryonic matter, the dark sector may host a variety of new, interacting particles that give rise to phenomena as rich and diverse as the astrophysical structures of luminous matter. In particular, if the dark matter can efficiently dissipate its kinetic energy, then dark matter itself can collapse to form compact objects such as dark black holes (DBH) \citep{damico_massive_2018,shandera_gravitational_2018,Chang:2018bgx,Choquette2019,latif_black_2019}, dark white dwarfs \citep{ryan2022exotic}, or dark neutron stars \citep{Hippert2021}.

The mass spectrum of such compact objects is directly related to dark-matter physics. Most strikingly, black holes below the baryonic Chandrasekhar limit of about a solar mass cannot form through standard stellar evolution. But, the Chandrasekhar limit $M_{\rm Chandra}\propto m_{\rm Planck}^3/m_{\rm proton}^2$ is determined by the mass of the proton and so the analogous limit could be much smaller for dark-matter fermions heavier than 1 GeV. Constraints on compact object mergers can therefore constrain the microphysics of dark matter \citep{singh_gravitational-wave_2020}.

Inferring the dark-matter microphysics from the observed mass spectrum, however, requires an accurate forward model of the DBH formation process. To make progress, we consider the scenario known as ``atomic dark matter'' (aDM) \citep{goldberg_new_1986,Ackerman2009,Feng2009,kaplan_atomic_2010, kaplan_dark_2011,cyr-racine_cosmology_2013,cyr-racine/etal:2014,Fan2013,Cline2014,boddy_hidden_2016,2015PhRvD..91b3512F,2016JCAP...07..013F,2015JCAP...09..057R,Agrawal2017,Ghalsasi:2017jna}, in which the dark matter consists of two fundamental fermions (one heavy and one light) oppositely charged under a dark electromagnetism. This model provides a realistic yet tractable venue in which to explore the consequences of dissipative dark matter. In this {\it Letter}, we present the fragmentation mass scale for atomic-dark-matter halos by solving the evolution of the thermo-chemical network, including both dark atomic and molecular cooling processes.

In the aDM model, dark molecular hydrogen provides the dominant cooling mechanism at temperatures below the dark-atomic-cooling limit.  This parallels the formation process of the first, or Population~III, stars which formed in pristine, low-metallicity gas clouds. There, the minimum temperature of the gas imprints a characteristic mass scale which determines the mass of the first stars \citep{Bromm_2002}. Previous work has used a simple criterion comparing cooling time from atomic processes to free-fall time to determine if a halo could cool \citep{buckley_collapsed_2018}, and estimated the DBH mass as a function of the dark-matter parameters by re-scaling results from the Pop.~III literature \citep{shandera_gravitational_2018}. In that estimate, the energy of the lowest allowed molecular transition was assumed to determine the coldest temperature reached by the gas. However, our recent work \citep{darkchem, darkkrome, darkrec} which derives the relevant molecular processes now allows a direct calculation of the cooling efficiency of the dark matter over a range of model parameters. These parameters are as follows: the mass of the heavy fermion, $M$; the mass of the light fermion, $m$; the fine structure constant, $\alpha_D$; the dark photon temperature, $T_D$. We refer also to these parameters by the ratios to their baryonic counterparts: $r_M = M/m_p$, $r_m = m/m_e$, $r_\alpha = \alpha_D/\alpha$, $\xi = T_D/T_{\rm CMB}$.
Here, $p$, $e$ and ${\rm CMB}$ stand for proton, electron, and the cosmic background photon.

If $\xi \sim 0.5$, at most $5\%$ of the dark matter can be atomic while satisfying the constraint from the lack of dark acoustic oscillations in the galaxy two-point correlation function \citep{cyr-racine_constraints_2014}. Such a high value of $\xi$ can be realized for dark matter thermally coupled to the Standard Model particles at an early time if there are no additional relativistic degrees of freedom in either the Standard Model or the dark sector \cite{agrawal_dark_2017}. Relaxing those assumptions allows smaller $\xi$ and opens up a parameter space where the dark matter can be entirely atomic. We do not know of a fundamental lower bound on $\xi$, but for $\xi \lesssim 10^{-3}$ the calculation of the thermal and chemical evolution of the primordial universe is complicated by the high dark particle to dark photon ratio, $\eta_D$ \citep{Gurian_2022}. A larger value of $\xi$ on the other hand increases the primordial free electron fraction, which initially promotes atomic cooling and molecule formation. However, evading the aforementioned constraints from large scale structure with a higher value of $\xi$ requires either a rather heavy dark electron  $r_m \gtrsim 10$ or a large coupling constant $r_\alpha >1$. Here we take $\xi = 5\times 10^{-3}$ for which the dark matter can be entirely atomic over a wide range of $r_m$ while $\eta_D$ is tractably small. We further simplify the parameter space by choosing $r_\alpha = 1$. The radiative cooling rates have strong dependencies on $r_\alpha$, so efficient cooling requires that $r_\alpha$ not be too much smaller than one. On the other hand, constraints on the dark matter self-scattering cross-section, which generally take the form of the upper bounds on $\sigma/M$, tend to prefer some combination $r_\alpha<1$ or $r_M>1$. For a more detailed discussion of the constraints on the dark matter cooling curve at a range of scales, see \citet{singh_gravitational-wave_2020}. With these two parameters fixed, we study the fragment mass for a range of $r_m$ and $r_M$.

The full dynamical evolution of a collapsing gas cloud requires three-dimensional simulations accounting for gravity and hydrodynamics. We instead compute the chemical and thermal evolution of a homogeneous (uniform density) parcel of gas. It turns out that these simple calculations have a clear correspondence with the results of full three-dimensional simulations \citep{Yoshida_2006}. As suggested by \citet{Glover2005} (Section 2.4) and \citet{Glover2008} (Eqs.~53-57), we bracket the likely range of results that full simulations may find by considering two extremes for the density evolution: adiabatic evolution corresponding to unrestricted gravitational runaway (pressure-free collapse) and pressure-supported hydrostatic evolution where the number density is held constant. For the latter, we assume a halo that cools efficiently with a constant density can collapse the core because the cooling will eventually eliminate the pressure support. Because the molecule formation rates and cooling rates both increase with density, the hydrostatic case represents the highest plausible minimum temperature. On the other hand, the adiabatic case allows the molecular fraction and cooling rates to attain their maxima.

The adiabatic collapse calculation is carried out using DarkKROME \citep{darkkrome}, while the hydrostatic calculation is implemented independently, closely following \citet{tegmark_how_1997}. The true thermo-chemical evolution of a three-dimensional self-gravitating gas cloud must lie somewhere between what we calculate under these extreme assumptions. Both calculations produce a minimum temperature for the collapsing gas, which is associated with the final mass of any compact object formed. We calculate a lower bound on this mass using the opacity limit argument of \citet{Low/Lynden-Bell:1976} and \citet{Rees76}.

\section{Thermo-chemical evolution}
\begin{figure*}
    \centering
    \includegraphics[width=\textwidth]{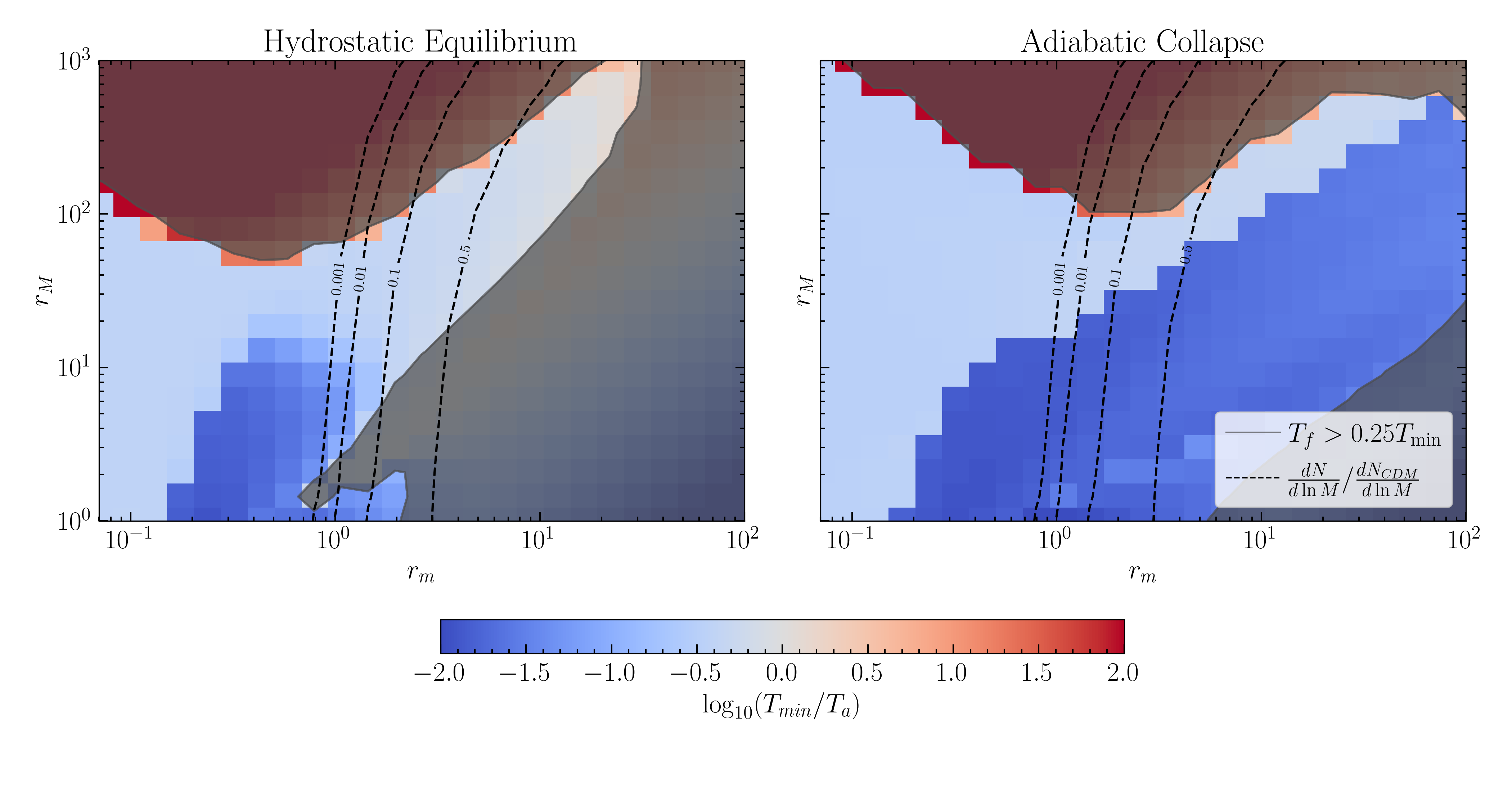}
    \caption{The minimum temperature of a $10^8 \, \rm M_\odot$ atomic dark matter halo at $z=10$ with $r_\alpha = 1$ and $\xi = 5\times 10^{-3}$, under constant-density evolution (left) and adiabatic collapse (right). The region which does not cool substantially is shaded in dark gray. Light blue/gray corresponds to cooling to the atomic limit, while dark blue indicates molecular cooling. Molecular cooling is more efficient under adiabatic collapse because the compressional heating and increasing density promote molecule formation. Contours of equal suppression of the halo mass function compared to CDM are shown (dashed): to the left of these contours CDM-like virialized halos are very rare.}
    \label{fig:tmin}
\end{figure*}
Using DarkKROME \citep{darkkrome}, we trace the evolution of the adiabatic collapse of a uniform-density element for a range of dark parameters. For a halo of mass $M_{\rm Halo}$ at redshift $z$, the calculation is initialized at the virial density $\rho_V = 178 \bar\rho_m(z)$ and virial temperature 
\begin{equation}
    T_V = \left(\frac{4\pi}{3}\rho_V\right)^{1/3}\frac{G r_M m_H}{5k_B} M_{\rm Halo}^{2/3},
    \label{eq:TV}
\end{equation} 
with $G$ the gravitational constant, $m_H$ the mass of Standard Model hydrogen, $k_B $ the Boltzmann constant and $M_{\rm Halo}$ the halo mass. Note that the halo mass enters our calculation only by determining the initial temperature of the gas parcel. The density evolves as
\begin{equation}
		\frac{d\rho}{dt}=\frac{\rho}{t_{ff}},
		\label{eq:ffevol}
	\end{equation}
	with the free-fall time $t_{ff}= \sqrt{(3\pi)/(32 G \rho)}$, and the temperature evolves according to the energy balance equation:
    \begin{equation}
    \label{eq:EvolveTemp}
	    	\frac{dT}{dt}=(\gamma-1)\frac{\Gamma(T,\{n_i\})-\Lambda(T,\{n_i\})}{k_B \sum_i n_i},\,
\end{equation}
where $\gamma$ is the adiabatic index, $\Gamma$ the heating rate, $\Lambda$ the cooling function, and $n_i$ the number density of each dark matter species. We denote the set of all $n_i$ by $\{n_i\}$.  The heating rate is dominated by compressional heating with $\Gamma_c= \sum_i n_i k_B T/t_{ff}$. The cooling rate $\Lambda$ includes both atomic and molecular processes, as discussed in \cite{darkchem, darkkrome}. We initialize the chemical abundances by the background abundances computed using the method described in \cite{darkrec}. We then simultaneously solve the chemical-reaction network of \cite{darkchem} and the thermal equation Eq.(\ref{eq:EvolveTemp}). Assuming secluded dark matter, we only compute for a dark-matter-only cloud, not including baryons. The calculation is halted either when the halo exceeds the virial temperature at the current density by a factor of 5 (indicating that ignoring pressure support is grossly inappropriate) or at the density threshold above which the collapsing cloud is opaque to molecular lines (Appendix~\ref{app}).

Independent of DarkKROME, we solve the same chemical-reaction network in hydrostatic equilibrium with a constant density, fixed at $\rho = \rho_V$, inspired by \cite{tegmark_how_1997}. The temperature is again evolved according to \refeq{EvolveTemp}. This implementation uses simpler molecular cooling rates, re-scaled from \cite{Hollenbach1979}, and neglects heating and cooling due to the formation and destruction of molecules (endo- and exo-ergic processes). At constant density, the compressional heating vanishes, and we evolve the network for a free-fall time at the virial density. We have checked that our analysis reproduces the results of \citet{tegmark_how_1997}, who tracked the density evolution through spherical collapse, and that the result agrees with the DarkKROME in the appropriate low-density limit. 

Collapse and fragmentation first occur within the dark-matter halos exceeding the Jeans mass. In the baryonic sector, such halos form at $z \approx 20$  with mass $M \sim 10^6 \,\rm M_\odot$ \citep{Glover_2012, stiavelli}. Atomic dark matter introduces modifications to the linear matter power spectrum through dark acoustic oscillations (DAO) and diffusion damping \citep{cyr-racine_cosmology_2013,darkrec}, both of which suppress the abundance of dark-matter halos with low mass. We therefore focus on somewhat heavier $10^8 \, \rm M_\odot$ halos that form later, around $z=10$. A suppression scale of the linear power spectrum around or below halo mass of $10^8 \rm M_\odot$ is consistent with the suppression scale of the allowed $\sim \rm keV$ warm dark matter \citep{Smith/Markovic:2011}. For both aDM and WDM bound structures which do not resemble virialized cold dark matter (CDM) halos may persist below this scale \citep{Angulo_2013, Stucker_2021}. Their role in the formation of compact objects requires further study.

In \reffig{tmin}, we show the minimum temperature obtained in our calculations for $10^8 \rm M_\odot$ halos at redshift 10 (when $10^8 \, M_\odot$ halos are common). The color bar shows the ratio of temperature to the atomic-limit temperature $T_a = 10^{4} r_m r_\alpha^2\, \rm K$. In the hydrostatic case (left panel), this minimum temperature is simply the temperature after one free-fall time since there is no heating. We have also shaded in gray the region which fails to cool by at least a factor of four from its initial temperature. This region fails to radiate its thermal energy within a free-fall time and likely does not undergo runaway gravitational collapse. Because $T_v \propto r_M^{2/3}$, in the bottom right portion of the figure cooling is actually prohibited by the low initial $T/T_a$. Although this region is ``cold'', it will remain near the initial, virial equilibrium rather than collapsing. In the adiabatic case (right panel), the minimum temperature is the lowest temperature achieved before the density cutoff due to high opacity (See Appendix \ref{app}). Typically, the gas begins to heat or evolve quasi-isothermally well before the end of the run. Those halos which fail to ever cool significantly below the initial temperature are unlikely to fragment down to small scales\textemdash these too are shaded out in gray.

Cooling below the atomic-limit temperature (blue area) requires the active participation of molecules. The formation of molecular hydrogen requires free electrons, which are cosmologically scarce for $\xi=5\times 10^{-3} \ll 1$, that we have adopted. Therefore, molecular cooling requires either a) achieving a temperature high enough to increase the ionization fraction followed by atomic cooling, or b) a large enough primordial molecular fraction and molecular cooling rate that the primordial molecules alone can cool the halo. In case a) the primordial molecules are dissociated and the problem becomes largely insensitive to the primordial abundances (and hence the value of $\xi$). The $r_M$ and $r_m$ dependence of the molecular cooling rate decreases the cooling efficiency in the high $r_M$, low $r_m$ regime. In the adiabatic case, compressional heating can raise an initially low gas temperature sufficiently to ionize the gas. Moreover, the increasing density also boosts both the molecular cooling and formation rates, leading to molecular cooling in a larger part of the parameter space. In particular, the critical transition point between low density limit cooling (where every collisional excitation leads to a radiative decay, rate $\propto n^2$) and local thermal equilibrium cooling (where collisional de-excitation contributes, rate $\propto n$) exhibits a strong direct dependence on $r_m$. For vibrational cooling, $n_{\rm crit} \propto r_m^{19/4}$, so that molecular cooling remains efficient at high densities for large $r_m$.

\reffig{tmin} also shows the contours of increasing levels of suppression in the aDM halo mass function relative to the CDM case: to the right of these contours, aDM halos are about as common as in CDM. A higher value of $\xi$ would push these contours to the right (narrowing the interesting parameter space) while a smaller value would move the contours to the left. We use the Press-Schechter formalism for this calculation as done in \cite{darkrec}, where the halo mass is defined from the linear density field smoothed with a sharp-k filter. This mass assignment is $M_{sk} = \frac{4\pi}{3}\bar\rho[c R_{sk}]^3$, where $R_{sk}$ is the filter radius and $c=2.7$ is a calibration factor from simulations. For $\xi \ll 1$, the effect which controls the suppression scale is dark diffusion damping. The diffusion scale $k_D$ is given by \citep{Zaldarriaga/Harari:1995}
\begin{equation}
    \frac{1}{k_D^2}
=
\int_0^{a_{\rm dec}}
\frac{da}{a^2H(a)}\frac{1}{6(1+R)n_{\re_\rD}\sigma_{\rm T,D}a}
\left[\frac{16}{15} + \frac{R^2}{1+R}\right], 
\end{equation}
with $a$ the scale factor, $\sigma_{\rm T,D}$ the dark Thomson cross section and $R(a)= \frac{3\bar{\rho}_{DM}}{4\bar{\rho}_{\gamma,D}}$. For large $R$, this scale is very close to the horizon size at dark-photon decoupling, which is controlled by the temperature ratio $\xi$ and the dark atomic binding energy $E_{\rH_\rD} =r_m r_\alpha^2E_{\rH}$ (where $E_{\rH}$ is the Standard Model binding energy). Diffusion damping imposes a Gaussian cutoff in the linear power spectrum $P(k) \rightarrow P(k) e^{-(k/k_D)^2}$. In the Press-Schechter treatment \citep{Press1974, darkrec}, this strongly suppresses the halo mass function below the cutoff scale.
\begin{figure*}
    \centering
    \includegraphics[width=\textwidth]{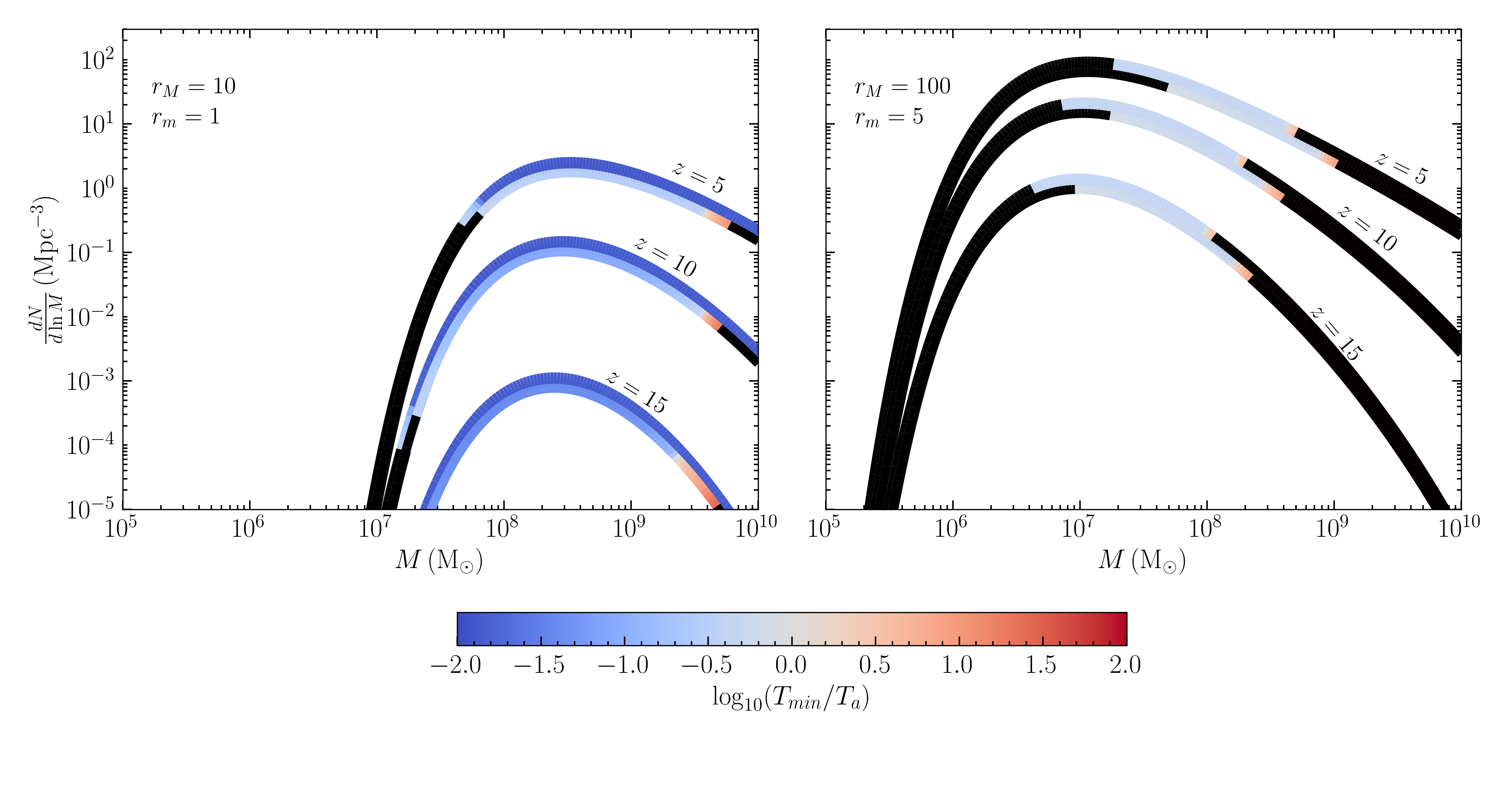}
    \caption{The halo mass function for two representative choices of parameters (see Fig.~\ref{fig:tmin}). The bottom color band of each curve indicates the minimum temperature of those halos which collapse under hydrostatic evolution, while the top band illustrates the adiabatic density case. The black region fails to cool. } 
    \label{fig:hmfs}
\end{figure*}

The dependence of these results on redshift and halo mass is shown in Fig.~\ref{fig:hmfs} by plotting the Press-Schechter halo mass function, with the range of halo masses that experience significant cooling (according to the virial temperature associated with the specified aDM parameters, redshift, and halo mass by \refeq{TV}) colored by their minimum temperature. As the density and virial temperature at fixed halo mass drop at low redshift, only increasingly hot (heavy) halos can cool. The halo mass function drops sharply below the cutoff scale. We emphasize again that the Press-Schechter formalism fails to capture any possible sheets and filaments below the cutoff scale \citep{Angulo_2013, Stucker_2021}.

\section{Minimum Mass}
\begin{figure*}
    \centering
   \includegraphics[width=\textwidth]{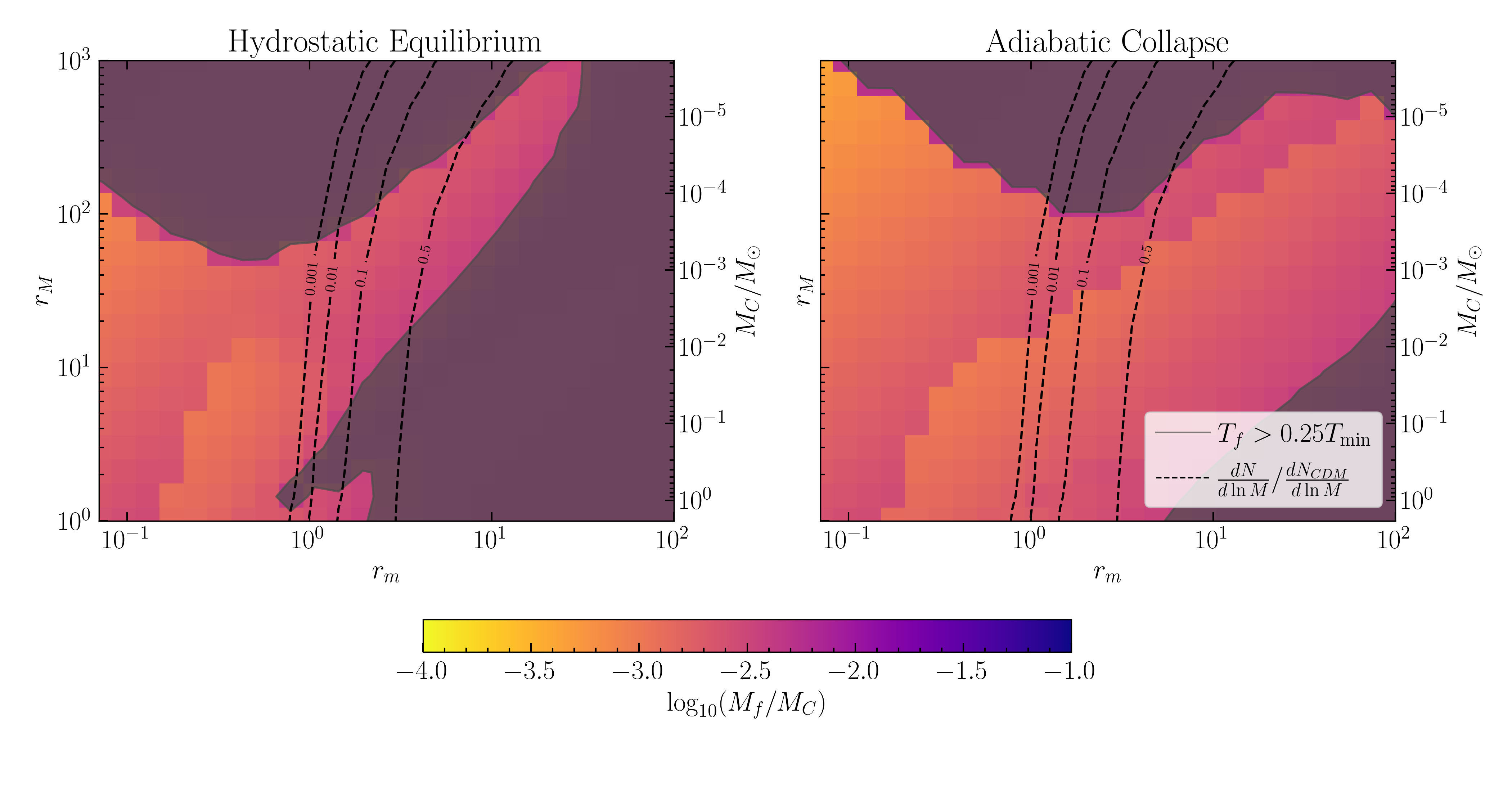}
    \caption{The opacity-limited minimum fragment mass for a $10^8 \, \rm M_\odot$ atomic dark matter halo at $z=10$ with $r_\alpha = 1$ and $\xi = 5\times 10^{-3}$ (as Fig.~\ref{fig:tmin}), under constant-density evolution (left) and adiabatic collapse (right). The region which fails to cool significantly (gray) and the halo-mass-suppression iso-contours (dashed) are again shown. The result is plotted as a fraction of the Chandrasekhar mass, $M_C$. The value of $M_C$ is shown on right-hand vertical axes.}
    \label{fig:results}
\end{figure*}
We can heuristically relate the minimum temperature to the mass of any compact objects based on the scaling of the Chandreskhar limit $M_{\rm C} \propto r_M^{-2}$ and the Jeans mass at fixed density $M_{\rm J} \propto T^{3/2}r_M^{-3/2}$, both of which imply that a large value of $r_M$ allows the formation of low-mass compact objects. A somewhat heavier dark proton, $r_M>1$, is generally preferred by constraints on the dark-matter momentum transfer, which provide upper bounds on $\sigma/M$ for the elastic scattering of neutral dark atoms \citep{Tulin/Yu:2018}. Of course, the temperature of the collapsing gas also depends on the dark matter parameters via the cooling rates. We can use the minimum temperature achieved in our calculations to compute a lower bound on the mass of any fragments in these halos using the opacity limit argument of \citet{Low/Lynden-Bell:1976} and \citet{Rees76}. In order for fragmentation to continue the fragment must be able to radiate ${\cal O}(1)$ of its gravitational binding energy within a free-fall time, while the radiation rate is bounded above by that of a black sphere. Equating these rates gives
\begin{align}
    M_f &\approx \left(\frac{\hbar c}{G}\right)^{3/2}\left(m_p r_M \right)^{-2}\left(\frac{k_B T}{r_M m_p c^2}\right)^{1/4} \nonumber \\
    & \approx M_{\rm Chandra} r_M^{-2} \left(\frac{k_B T}{r_M m_p c^2}\right)^{1/4}
    \equiv M_{C} \left(\frac{k_B T}{r_M m_p c^2}\right)^{1/4},
\end{align}
where  $M_C \equiv M_{\rm Chandra}r_M^{-2}$ is the Chandrasekhar mass of the dark matter \citep{shandera_gravitational_2018}. For baryons, $M_f \sim 0.045 f^{-1/2}(T/3000\,{\rm K})^{1/4}\, M_{\odot}$ with $f<1$ being the radiation efficiency \citep{Becerra/etal:2018}. We bound the analogous fragment mass in the aDM by inserting the minimum temperatures obtained above in this expression. We show the result $M_f/M_C$ in Fig.~\ref{fig:results}, where we have calculated the Chandrasekhar mass as $M_C \approx\left(\frac{\hbar c}{G}\right)^{3/2}\left(m_p r_M \right)^{-2} $. Note that for all realistic halos, $k_BT\ll r_Mm_p c^2$: the temperature is much smaller than the proton mass and the fragment mass is much smaller than the Chandrasekhar mass.

As a lower bound, this result is quite robust. However, caution is warranted in correlating this lower bound with the true mass of compact objects in the halo. If accretion is inefficient, these fragments may remain below the Chandrasekhar limit (where fermionic degeneracy pressure can halt collapse). On the other hand, for Pop.~III stars, the eventual stellar mass (after accretion) exceeds the opacity limit by a factor of $\sim 10^4-10^5$. For atomic dark matter fragments, such growth would imply the eventual formation of black holes, which could still be easily sub-solar-mass. Of course, this growth factor depends on the detailed evolution of the collapsing gas cloud and subsequent star-forming disk \citep{Hirano/etal:2014}. As such, our result cannot be directly translated to a final (post-accretion) mass for the compact objects.  Instead, Fig.~\ref{fig:results} should be taken as a qualitative measure of the instability of the halo to small-scale fragmentation. In this light, our calculations indicate that a substantial portion of the parameter space is dramatically more prone to forming low-mass fragments as compared to baryonic matter. 

\section{Summary and conclusion}
We have modeled the thermo-chemical evolution of atomic dark matter halos by following the evolution of an infinitesimal volume element, including both atomic and molecular processes. The density evolution is independently specified, and we study the results in two extreme cases: adiabatic collapse ($\dot{\rho}=\rho/t_{ff}$) and hydrostatic equilibrium (constant density), to study the range of possibilities. We also illustrate the dependence of these results on halo mass and redshift. The mass scale of the seed dark-matter fragments forming in the halos is determined by the minimum temperature the fragmenting gas reaches during its collapse. As an early exploration of the dependence of this mass scale on the model parameters, we have calculated the opacity-limited minimum fragment mass at this minimum temperature. We show that there is a substantial parameter space where these halos can fragment on scales far below one solar mass. Moreover, the Chandrasekhar mass of the dark matter is ${\mathcal{O}(1\, \rm M_\odot)}/ r_M^{2} $, which can clearly also be much less than one solar mass. Unless accretion is dramatically enhanced compared to the baryonic case, these halos may host compact objects that, at formation time, are orders of magnitude smaller than allowed by astrophysical processes in baryonic matter. We leave the final size of the compact objects, which requires modeling of accretion physics, as an object of future work. 

\acknowledgements

We thank Neal Dalal and Daniel Egana-Ugrinovic for discussions about the eventual fate of gas fragments. Funding for this work was provided by the Charles E. Kaufman Foundation of the Pittsburgh Foundation. This work was supported at Pennsylvania State University by NASA ATP Program No. 80NSSC22K0819. DJ is also supported by KIAS Individual Grant PG088301 at Korea Institute for Advanced Study.

\bibliographystyle{aasjournal}
\bibliography{darkfragments.bib}
\appendix
\section{Molecular Line Opacity}\label{app}
Throughout, we have assumed optically thin cooling, where every emitted photon escapes the collapsing gas cloud. This assumption breaks down at higher densities where cooling is inefficient due to the high opacity. Therefore, the minimum temperature is achieved before opacity becomes significant, so we stop the computation at the density where the opacity becomes large. Since the atomic cooling is fairly insensitive to the stopping condition, we consider only the molecular line opacity:
\begin{equation}
    \alpha_{\nu} = \frac{h \nu}{4\pi} \phi(\nu) (n_u B_{u\ell} - n_\ell B_{\ell u}),
\end{equation}
where $\nu$ is the photon frequency, $\phi{(\nu)}$ is the line profile function, $n_u$ is the number density in the upper state, $n_\ell$ is the number density in the lower state, and $B_{u\ell}$ and $B_{\ell u}$ are the Einstein coefficients. 
For this order-of-magnitude estimate, we can ignore the second term that represents stimulated emission. The $B$ coefficient is related to the $A$ coefficient as 
\begin{equation}
    B_{u\ell} = \frac{c^2}{2h\nu_0^3} A_{u\ell},
\end{equation}
with $\nu_0$ the line center.
Then, 
\begin{equation}
    \alpha_{\nu} = \frac{c^2 \nu}{8\pi\nu_0^3} \phi(\nu) n_u A_{\ell u}.
\end{equation}
To calculate the photon escape probability, we evaluate the Sobolev optical depth \citep{Sobolev1960,Yoshida_2006, seager_how_2000}: 
\begin{equation}
    \tau = \frac{\alpha_{\nu_0}/\phi(\nu_0)}{\Delta \nu}L.
\end{equation}
where $\Delta\nu = \frac{\nu_0}{c} v_{\rm thermal}$ the thermally broadened line width, and $L$ the Sobolev length, which is the distance at which the velocity gradient in the collapsing cloud Doppler-shifts the photon out of the line. This length scale is 
\begin{equation}
    L = \frac{v_{\rm thermal}}{dV_r/dr} \approx v_{\rm thermal} t_{ff},
\end{equation}
where is $V_r$ the radial velocity gradient and $t_{ff}$ the freefall time. 
Then, inserting $t_{ff}$ and noting that $v_{\rm thermal}$ cancels,
\begin{equation}
    \tau = \frac{c^3 n_{u} A_{\ell u}}{8 \pi \nu_0^3}\sqrt{\frac{3\pi}{32 G r_M m_H n_{\rm tot}}}.
\end{equation}
If the excited states are in thermal equilibrium when opacity begins to contribute and a few excited states are populated, (as is the case for Pop.~III stars) then $n_u$ for each populated state is a factor a few smaller than $n_{\rm tot}$. We neglect this factor by equating $n_{\rm tot}$ and $n_u$, $n_u=n_{tot}\equiv n$, for 
\begin{equation}
   n= \frac{2048 \pi \nu^6 G r_M m_H\tau ^2}{3c^6 A_{\ell u}^2}.
\end{equation}
By considering the order of magnitude of the relevant Einstein coefficients in \cite{Turner77} and inserting the scaling of $\nu_0$ and $A_{u \ell}$ with the dark parameters from \cite{darkchem}, we have 
\begin{align}
    n_{\rm rot} &\approx \left[r_{\alpha}^{-2}r_M^5\right]10^{8}\tau^2\,{\rm cm^{-3}}\\
    n_{\rm vib} &\approx \left[r_{\alpha}^{-2}r_m^{2}r_M^{3}\right]10^{6}\tau^2\,{\rm cm^{-3}}.
\end{align}
For Pop.~III stars, this opacity begins to contribute at densities around $10^{10} \, \rm cm^{-3}$ and temperatures slightly less than $1000 $ K, before saturating around $10^{15} \, \rm cm^{-3}$. Noting that the Sobolev escape fraction $(1-\exp(-\tau))/\tau$ for $\tau = 1$ is 63\% while for $\tau = 10$ it is 10\% and that at low temperatures the rotational transitions dominate the cooling, taking $r_m = r_M =r_\alpha = 1$ is roughly consistent with the threshold $n \sim 10^{10} \, \rm cm^{-3}$ found in the literature. 

Above this density threshold, an adiabatically collapsing gas cloud cooled by molecular transitions will begin to heat. That is, the minimum temperature of our adiabatic runs is achieved before $n_\tau \equiv n_{\rm rot}$. Similarly, the transition at $n_{\rm crit}$ from low-density ($\propto n^2$) to high-density/local-thermodynamic-equilibrium (LTE, $\propto n$) molecular cooling will cause such a cloud to heat, as the LTE cooling is outcompeted by adiabatic heating ($n^{3/2}$). Therefore, the simulations must run at least until the particle density crosses the $\min(n_{crit},n_\tau)$ threshold. In contrast, we can only trust the computation up to $n_\tau$, since our cooling rates are valid both in the LDL and in LTE but not in the optically thick regime. Therefore, we stop the calculation at $n_{f}=\min(n_{\tau}, n_{\rm crit})$.

\end{document}